\PassOptionsToPackage{dvipsnames}{xcolor}
\documentclass[aps,prb,notitlepage,twocolumn]{revtex4-2}
\usepackage{amsmath}
\usepackage{graphicx}
\usepackage{slashed}
\usepackage{bm}
\usepackage{mystyleold} 
\newcommand{\adb}{\allowdisplaybreaks } 
\newcommand{\ann}{\adb \nonumber \\}
\newcommand{\ii}{\mathrm{i}}
\newcommand{\kB}{\bm{k}}
\newcommand{\uB}{\bm{u}}
\newcommand{\vB}{\bm{v}}
\newcommand{\sigmaB}{\bm{\sigma}}
\newcommand{\etaB}{\bm{\eta}}
\newcommand{\varepsilonB}{\bm{\varepsilon}}
\newcommand{\PiB}{\mathbf{\Pi}}
\newcommand{\tm}{\mathrm{tm}}
\newcommand{\te}{\mathrm{te}}
\newcommand{\id}{\mathrm{id}}
\newcommand{\m}{\mathrm{m}}
\newcommand{\GB}{\mathbf{G}}
\newcommand{\AC}{\mathcal{A}}
\newcommand{\RC}{\mathcal{R}}
\newcommand{\dd}{\mathrm{d}}
\newcommand{\gs}{\mathrm{g}}
\newcommand{\gr}{\mathrm{gr}}
\newcommand{\etaBT}{\widetilde{\etaB}}
\newcommand{\PhiT}{\widetilde{\Phi}}
\newcommand{\kT}{\tilde{k}}
\newcommand{\PiT}{\widetilde{\Pi}}
\newcommand{\IB}{\mathbf{I}}
\newcommand{\TB}{\mathbf{T}}
\newcommand{\DB}{\mathbf{D}}
\DeclareMathOperator{\diag}{diag}
\DeclareMathOperator{\sign}{sign}
\DeclareMathOperator{\tr}{tr}
\DeclareMathOperator{\arctanh}{arctanh}
\DeclareMathOperator{\Res}{Res}
\newcommand{\us}[2]{\underset{#2}{#1}}
\usepackage{xcolor} \definecolor{darkgreen}{rgb}{0,.5,0} \definecolor{background}{HTML}{FFFECF}
\usepackage[colorlinks,filecolor=blue,citecolor=darkgreen,unicode]{hyperref}
\begin{document}
%--------------------------------------------------------------------
\title{Casimir-Lifshitz force for moving graphene}
	\author{Mauro Antezza}	\email{mauro.antezza@umontpellier.fr}
	\affiliation{ Laboratoire Charles Coulomb (L2C), UMR 5221 CNRS-University of Montpellier, F-34095 Montpellier, France}
	\affiliation{Institut Universitaire de France, 1 rue Descartes, Paris Cedex 05 F-75231, France}
	\author{Natalia Emelianova}	\email{natalia7emelianova@gmail.com}
	\author{Nail Khusnutdinov}\email{nail.khusnutdinov@gmail.com}
	\affiliation{CMCC, Universidade Federal do ABC, Avenida dos Estados 5001, CEP 09210-580, SP, Brazil}
	\date{\today}
	\begin{abstract}
		We consider the system of two parallel sheets of graphene which are moving with relative parallel velocity $\vB$ and calculate the Casimir energy by using the scattering approach. The energy has real and imaginary parts. The real part contains a velocity contribution to the force perpendicular to the sheets. The imaginary part is related to a friction force parallel to the graphene sheets. We prove that the friction appears if the relative velocity becomes greater than the velocity of Fermi and if the modulo of reflection coefficient is greater than the one which is significant for virtual photon production. We analyze in detail the real contribution to the Casimir energy for two systems -- graphene/graphene and ideal metal/graphene. In the non-relativistic case $v \ll v_F$, the relative correction to the Casimir energy $(\mathcal{E}_v - \mathcal{E}_0)/\mathcal{E}_0$ is proportional to the $(v/c)^2$ (the maximum value is $0.0033$ for the gapeless case and $v=v_F$) for the first system, and it is zero up to the Fermi velocity $v = v_F$ for system ideal metal/graphene.
	\end{abstract}  
	\maketitle 

\section{Introduction} 

Nowadays, the Casimir effect \cite{Casimir:1948:otabtpcp,*Lifshitz:1956:tmafbs,*Dzyaloshinskii:1961:gtvwf} is a well-known and experimentally observed phenomenon -- the macroscopic evidence of the microscopic quantum origin of fields and matter (see, for example, recent review \cite{Woods:2016:MpoCavdWi} and book \cite{Bordag:2009:ACE}). The Casimir effect for real materials depends on the physical properties of solids and shapes of their boundaries, temperatures of bodies and fields, chemical potential, etc. \cite{Bordag:2009:ACE}. The different temperatures of the solids lead to heat transfer, and the non-equilibrium Casimir effect \cite{Antezza:2004:eotcpfotcooatbec,*Antezza:2005:nabotsafoote,*Pitaevskii:2006:LbsCfote,*Obrecht:2007:mottdotcpf,*Antezza:2008:clfoote} takes place. The new materials such as graphene \cite{Klimchitskaya:2022:cegset,*Abbas:2017:staemotcfigm}, Weyl semi-metals \cite{Wilson:2015:rcfbws,*Rodriguez-Lopez:2020:scorCitIaIIWs,*Farias:2020:cfbwscm}, and topological matter \cite{Lu:2021:cetm}  demonstrate the great diversity of properties of the Casimir effect.  

Graphene is a mono-atomic two-dimensional layer of Carbon atoms,  and the main role of the Casimir effect for graphene plays its $2D$ conductivity tensor. In the frameworks of Kubo approach it was calculated in Ref. \cite{Gusynin:2006:tdqghoc} and by using $(2+1)$D quantum electrodynamics in Ref. \cite{Bordag:2009:CibapcagdbtDm,*Fialkovsky:2011:FCefg}. 
%The polarization tensor approach used in the last reference is more fundamental because it does not contain any free parameter such as, for example, scattering rate. 

The Casimir force depends on the relative velocity of slabs, too.  The uniform motion perpendicular to the slabs was considered in Ref. \cite{Bordag:1984:ccesfsnbc,*Bordag:1986:ceumm} for scalar field and electromagnetic fields (see also the recent review \cite{Dodonov:2020:fydce} on the dynamical Casimir effect). The relative movement with velocity parallel to the slabs gives two different contributions to the Casimir force -- perpendicular and parallel to the slabs. The first one, for dielectric layers, was considered in Ref. \cite{Maslovski:2011:crmm}, where repulsive Casimir force was observed. The second contribution is called van der Waals/Casimir (quantum or non-contact) friction. It was investigated for a long time (see, for example, \cite{Pendry:1997:svqf,*Philbin:2009:nqfbump,*Brevik:2022:feaoe} and reference therein and the recent review \cite{Volokitin:2007:nfrhtnf}).

The Casimir friction for graphene was considered in Ref.\,\cite{Farias:2017:QFbGS,*Farias:2015:fatqfeaadf} in the framework of effective action. It was obtained that the friction appears as an imaginal part of effective action and the threshold for the friction force appears -- it is zero if the relative velocity is smaller than the velocity Fermi. This statement is in agreement with Ref. \cite{Maghrebi:2013:qcranf} where it was shown that a frictional force arises when the relative velocity exceeds that of light in the medium. The Dirac electron is described by the Dirac equation with velocity Fermi instead of the speed of light.

In this paper, we use the following approach to obtain the Casimir energy of graphene moving with relative parallel velocity. By using the Lorentz transformation of graphene's conductivity tensor \cite{Bordag:2009:CibapcagdbtDm}  we obtain the tensor of the moving graphene and then use the formulas obtained in Ref. \cite{Fialkovsky:2018:qfcrbcs} to obtain the real and imaginary parts of the Casimir energy and the force. To do this we rotate the contour of integration from real frequencies to imaginary ones. The integration over the imaginary axis gives the real contribution to the Casimir force perpendicular to the sheets of graphene. The specific complex frequencies produce an imaginary contribution which is the Casimir friction -- the force parallel to the sheets.  In this paper, we consider the real contribution, only. 

The remainder of the paper is organized as follows. In Sec.\,\ref{Sec:General} we consider the Lorentz transformation for the action. In Sec.\,\ref{Sec:conductivity} the tensor of conductivity of moving graphene in the laboratory frame is derived. We use the definition of conductivity as a boundary condition in the form of Ohm low and the Lorentz transformation to the laboratory frame. 

The Sec.\,\ref{Sec:Casimir} is devoted to the derivation of the Casimir energy and force. We start from the expression for the energy at the real frequencies and then move the integration to the imaginary axes. Two parts of the energy appeared. The part of the energy with integration along the imaginary frequencies $\xi = -\ii \omega$ (Sec.\,\ref{Sec:Casimir2}) gives the \textit{real} contribution, only to the energy. This statement is the consequence of the general relations for the conductivity tensor. These relations are proved in the Appendix \ref{Sec:anprop}. The \textit{imaginary} contribution to the energy and force may come from the last part, only   (Sec.\,\ref{Sec:Casimir1}) which has the form of contour integrals along close contours in the upper and down half-planes of complex frequencies $\omega$ which corresponds to the opposite direction of scattering. The non-zero contributions from these integrals exist if some specific condition is satisfied. This condition in the scalar form has been noted by Lifshitz in Ref.\,\cite{Lifshitz:1956:tmafbs}  and also in Ref.\,\cite{Kampen:1968:otmtovdwf,*Gerlach:1971:eovdwfbsatspi}). 
In this section, we proved the threshold for velocity -- the imaginary contribution to the energy is zero up to the velocity of Fermi. We proved also that this contribution takes pales if and only if the modulo of the reflections coefficient is greater than 1, that is the photon production has to appear. 

In Sec.\,\ref{Sec:Real} we consider in detail the velocity correction to the perpendicular force, due to the real part of the energy, and for two systems -- graphene/graphene and graphene/ideal metal cases.  

The direct calculations of the boosted polarization tensor are shown in Appendix \ref{Sec:App1}.

We use the natural units $\hbar = c =1$ and the  following notations for indexes: $\mu,\nu = 0,1,2,3 = t,x,y,z$; $a,b = 0,1,2 = t,x,y$, and $i,j = 1,2 = x,y$. The wave vector, $k^a = (\omega, \kB)$, and $k'^a = (\omega', \kB')$.  

\section{General remarks}\label{Sec:General}

The presence of the velocity of Fermi instead of the velocity of light in the $3$D Dirac equation breaks the Lorentz invariance of the action. Indeed, in the laboratory frame, $K$ the action for a single graphene sheet reads 
\begin{equation}
	\AC = - \frac{1}{4} \int  \dd^4x F^2 + \int \dd^3 x \bar{\psi} \left[\tilde{\gamma}^b (\ii \partial_b - e A_b) -m \right]\psi |_{z=a},
\end{equation}
where $\tilde{\gamma}^a = \tilde{\gs}^a_b \gamma^b$ and $\tilde{\gs}^a_b = \diag (1,v_F,v_F)$. The graphene is situated perpendicular to axis $z$ at point $z=a$. Then we make a boost in the plane of graphene with velocity $\vB$ to the moving frame $K'$: $x'^a = \Lambda^{a'}_{\cdot b} x^b$, where the Lorentz matrix reads
\begin{equation}%\label{eq:Lorentz}
	\Lambda = [\Lambda^{a'}_{\cdot b}] = 
	\begin{pmatrix}
		\gamma & - \gamma\vB\\
		-\gamma\vB & \IB + (\gamma -1)\frac{\vB \otimes \vB}{\vB^2}
	\end{pmatrix},
\end{equation}
and $\gamma = 1/\sqrt{1-\vB^2}$. The spinor transforms with matrix $S = \exp(-\frac{\ii}{4} \sigma^{ab} \lambda_{ab})$: $\psi'(x') = S \psi(x)$, $\bar{\psi}'(x') = \bar{\psi}(x) S^{-1}$, where 
\begin{equation}
	\sigma^{ab} = \frac{\ii}{2}[\gamma^a,\gamma^b],\ \lambda_{ab} = 
	\begin{pmatrix}
		0&-v^1 &-v^2\\
		v^1 &0&0\\
		v^2 & 0& 0 	
	\end{pmatrix}.
\end{equation} 

The action in the frame $K'$ becomes
\begin{equation*}
	\AC' = - \frac{1}{4} \int  \dd^4x' F'^2 + \int \dd^3 x' \bar{\psi}' \left[\tilde{\Gamma}^b (\ii \partial_{b'} - e A'_b) -m \right]\psi'|_{z=a},
\end{equation*}
where 
\begin{eqnarray}
	\tilde{\Gamma}^{a'} &=& \Lambda^{a'}_{\cdot b} S\tilde{\gamma}^b S^{-1}  = \Lambda^{a'}_{\cdot b}  \Lambda^c_{\cdot e'} \tilde{\gs}^b_c\gamma^{e'} = \tilde{\gs}'^{a'}_{e'} \gamma^{e'} \ann
	&=& v_F \gamma^a + u^a (1-v_F) \slashed{u},
\end{eqnarray}
and $\slashed{u} = \gamma^a u_a$, $u_a = (u_0,u_1,u_2) = \gamma (1,-\vB)$. Here the matrix $\tilde{\gs}' = \Lambda^{-1}\tilde{\gs}\Lambda$ is the Lorentz transform of $\tilde{\gs}$. We observe that $\tilde{\Gamma}^{a'} \not= \tilde{\gamma}^{a'}$ if $\vB \not = \bm{0}$, and therefore, $\AC' \not = \AC$. For $v_F=1$ we obtain the Lorentz invariance, $\tilde{\Gamma}^{a'} = \tilde{\gamma}^{a'} = \gamma^{a'}$, as should be the case. 

\section{The conductivity of moving graphene}\label{Sec:conductivity}

Let us consider a system of two parallel graphenes which are situated perpendicular to the axis $z$ and are separated by distance $a$ apart. The first graphene is at rest in the laboratory frame $K$ and the second graphene is moving parallel to the first graphene with the $3$-velocity $u^a = (u^0,\uB)$. The first graphene is described by the tensor conductivity $\sigmaB_1$ and the second one -- by the tensor conductivity $\sigmaB_2$. Both conductivities are for the laboratory frame $K$.  

The expression for tensor conductivity of graphene has been calculated in the co-moving frame $K'$ in Ref. \cite{Bordag:2009:CibapcagdbtDm,*Fialkovsky:2011:FCefg} for zero temperature and non-zero temperatures in terms of the polarization tensor in $3$D quantum electrodynamics. To obtain the tensor conductivity of graphene $\sigmaB_2$ in the laboratory frame $K$ we make the Lorentz boost from the co-moving frame $K'$ to the laboratory frame $K$. The conductivity of the second graphene in its co-moving frame one denotes with prime, $\sigmaB'_2$. 

Let us repeat the derivation of this tensor and extend it to the problem under consideration. The boundary conditions in the co-moving frame $K'$, where graphene is at rest,  have the following form 
\begin{align}
	[A'_{\mu}(x')]_{z' = a} & =  0,\nonumber \\{}
	[F'^{z \nu}(x')]_{z' = a} & =  \int \dd^3  \tilde{x}'' \PiT'^{\nu \rho}  (x' - x'') A'_{\rho} (x'') |_{z' = z'' = a}, \label{eq:BC-1}
\end{align}
where $[f]_{z'=a} = f_{z'=a-0}-f_{z'=a+0}$  and $\PiT'^{3 \mu} = \PiT'^{\mu 3} = 0$. In what follows, we will assume that $z = z' = z'' = a$. Then, we make the Fourier transform with respect to the three coordinates $x'^a$ and obtain  
\begin{align}
	[A'_{\mu}(k')] & =  0,\nonumber \\{}
	[F'^{z \nu} (k')] &= \PiT'^{\nu \rho} (k') A'_{\rho} (k'). \label{eq:BC-2}
\end{align}
Let us consider first the case $\nu=i$. By definition $F'_{i 0} = \partial_{i'} A'_0 - \partial_{0'} A'_i = - E'_i(x')$  and therefore $A'_i = E'_i / \ii \omega' + A'_0 k_i / \omega'$. Then, by using the gauge invariance condition $\PiT'^{\nu \rho} (k')k'_\rho =0$, the second boundary condition \eqref{eq:BC-2} is represented in the following form  
\begin{equation}
	[F'^{z i}(k')] = \frac{1}{\ii \omega'} \PiT'^{ij} E'_{j}.
\end{equation}
It has the form of boundary condition for tangent components of a magnetic field 
\begin{equation}
	[\vec{H}'] \times \vec{n} = \vec{J}', 
\end{equation}
with surface current 
\begin{equation}
	J'^{i}= \sigma'^{i j} E'_{j},
\end{equation}
and normal vector $n^a = \delta^a_z$. Therefore, 
\begin{equation} \label{eq:PT-1}
	\sigma'^{ij} = \frac{1}{\ii \omega'} \PiT'^{ij}.
\end{equation}

By setting now the index $\nu = 0$ in Eq.\,\eqref{eq:BC-2} we obtain the boundary condition for the normal component of the electric field 
\begin{equation}
	[\vec{E}']\cdot \vec{n} =  -\frac{1}{\ii \omega'} \PiT'^{0j} E'_{j} = -\rho'.
\end{equation}

To obtain the conductivity tensor, $\sigmaB$, in the laboratory frame $K$, we make boost, $K' \to K:$ $x'^a = \Lambda^{a'}_{\cdot b} x^b$, $k'_a = \Lambda_{a'}^{\cdot b} k_b$, in the boundary equations \eqref{eq:BC-2} to the laboratory frame with  matrix 
\begin{equation}\label{eq:Lorentz}
	\Lambda = [\Lambda^{a'}_{\cdot b}] = 
	\begin{pmatrix}
		u^0 & - \uB\\
		-\uB & \IB + \frac{\uB \otimes \uB}{u^0 +1}
	\end{pmatrix},
\end{equation}
where $\uB = (u^1,u^2) = \vB \gamma$ and $\IB$ is $2\times 2$ identical matrix.  In matrix notations, $x' = \Lambda x$ and $k' = k \Lambda^{-1}$. Then,  in the laboratory frame $K$ we obtain the boundary conditions 
\begin{align}
	[A_{\mu}(k)] & =  0,\nonumber \\{}
	[F^{z \nu} (k)] &=  \PiT^{\nu \rho} (k)A_{\rho} (k), \label{eq:BC-3}
\end{align}
where 
\begin{equation}\label{eq:PT-2}
	\PiT^{\nu \rho} (k) = \Lambda^\nu_{\cdot \nu'}\Lambda^\rho_{\cdot \rho'} \PiT'^{\nu' \rho'} (k').
\end{equation}
The same expressions for $\PiT^{\nu \rho} (k)$ may be obtained by direct calculation of boosted Dirac equations (see Appendix \ref{Sec:App1}).

By doing the same steps \eqref{eq:BC-2}-\eqref{eq:PT-1} but with wave vector $k$ instead of $k'$, we obtain the conductivity tensor of graphene in the laboratory frame $K$
\begin{equation}
	\sigma^{ij}(k) = \frac{1}{\ii \omega} \PiT^{ij}(k)=\frac{1}{\ii \omega}  \Lambda^i_{\cdot \nu'}\Lambda^j_{\cdot \rho'} \PiT'^{\nu' \rho'} (k').
\end{equation}

The PT has two contributions namely, symmetric (parity-even) and antisymmetric (parity-odd) $\PiT'^{\mu\nu} = \PiT'^{\mu\nu}_s + \PiT'^{\mu\nu}_a$,  where $\PiT'^{\mu\nu}_a = \xi\epsilon^{\mu\nu\alpha} k'_\alpha$. Due to gauge invariance $\PiT'^{\mu\nu} k'_\mu = \PiT'^{\mu\nu}_s k'_\mu \equiv 0$. We use these relations to express time components 
\begin{equation}\label{eq:Gauge}
	\PiT'^{00}_s  =  \frac{1}{\omega'^2} \PiT'^{ij}_s k'_{i}k'_{j},\ \PiT'^{0i}_s  =  -\frac{1}{\omega'} \PiT'^{ij}_s k'_{j}.
\end{equation}
Taking these relations into account we obtain
\begin{equation}
	\PiT^{ij} (k)  = \PiT'^{n'm'} G^i_{\cdot n'} G^j_{\cdot m'},
\end{equation}
where 
\begin{equation}
	G^i_{\cdot n'} = \delta^i_{n'} - \frac{u^i u_n}{u_0 + 1} - \frac{u^i k'_n}{\omega'}.
\end{equation}
By using manifest form of $k'= k \Lambda^{-1}$,
\begin{equation}\label{eq:kL}
	\omega' = (ku), \kB' = \kB - \uB \frac{\omega + (ku)}{1+u_0},
\end{equation} 
we obtain 
\begin{equation}
	G^i_{\cdot n'} =\delta^i_{n'}  +\frac{\omega u^i u_n}{(ku) (u_0+1)} - \frac{u^i k_n}{(ku)},
\end{equation}
where $(ku) = k^a u_a$. In the matrix notations, one has
\begin{equation}
	\GB =\IB  - \frac{\omega}{(ku)} \frac{\uB\otimes \uB}{u_0+1} + \frac{\uB \otimes \kB}{(ku)},
\end{equation}
and
\begin{equation}
	\tilde{\PiB}=  \GB \tilde{\PiB}' \GB^T.
\end{equation}

Therefore, the conductivity tensor $\etaB_i = 2\pi \sigmaB_i$ in the laboratory frame is 
\begin{equation}\label{eq:eta2}
	\etaB _2 = \frac{\omega'}{\omega} \GB\etaB'_2 \GB^T,
\end{equation}
where $\etaB'_2$ is the conductivity tensor of the second plane in its frame, which depends on the $\omega',\kB'$. The straightforward calculations give 
\begin{equation}
	\det \mathbf{G} = \frac{\omega}{\omega'} = \frac{\omega}{(k u)},
\end{equation}
and therefore $\det\etaB_2 = \det \etaB'_2$. 

A graphene at rest has the following conductivity \cite{Bordag:2009:CibapcagdbtDm} 
\begin{equation}\label{eq:eta1}
	\etaB_1 = \eta_{\gr} \frac{\kT}{\omega} \left(\IB + v_F^2 \frac{\kB \otimes \kB}{\kT^2}\right) \PhiT \left(\frac{\kT}{2m}\right),
\end{equation}
where $\eta_\gr = 2\pi \sigma_\gr = \pi \alpha/2 = 0.0114$, and 
\begin{align*}
	\PhiT(y) &= \frac{2\ii }{\pi y}\left\{ 1 - \frac{y^2 + 1}{y}\arctanh y\right\},\\
	\kT &=  \sqrt{\omega^2 - v_F^2 \kB^2}. 
\end{align*}
The conductivity  of the moving graphene has form \eqref{eq:eta2} where 
\begin{equation}\label{eq:etaP2}
	\etaB'_2 = \eta_{\gr} \frac{\kT'}{\omega'} \left(\IB + v_F^2 \frac{\kB' \otimes \kB'}{\kT'^2}\right)\PhiT \left(\frac{\kT'}{2m}\right).
\end{equation}
Here $\kT' = \sqrt{\omega'^2 - v_F^2 \kB'^2}$, or by using the Lorentz boost \eqref{eq:kL} one obtains 
\begin{equation}
	\kT' = \sqrt{\kT^2 + \frac{1-v_F^2}{1-v^2} (\omega^2 \vB^2 + (\kB\vB)^2 -2 \omega (\kB\vB) )}.
\end{equation}

The tensor \eqref{eq:eta2} is symmetric and depends on the vectors $\kB$ and $\vB$. Therefore, the general structure of this tensor is the following 
\begin{equation}
	\etaB_2 = A_1 \IB + A_2 \kB\otimes\kB + A_3 \vB\otimes\vB + A_4 (\kB\otimes\vB + \vB\otimes\kB).
\end{equation}
Calculating different contractions of the above relation with $\IB$, $\kB$, and $\vB$ we obtain 
\begin{equation}
	\etaB_2 = i_1 \IB  + i_2 \kB\otimes\kB + i_3 (\kB\otimes\vB + \vB\otimes\kB),
\end{equation}
where
\begin{align}
	i_1 &= \frac{\eta _\gr \PhiT'}{\omega  \kB^2 \kT'}  \left(\kB^2 \kT'^2 + \frac{1-v_F^2}{1-v^2} ((\kB\vB)^2 - \kB^2 \vB^2)k_3^2\right),\ann
	i_2 &= \frac{\eta _\gr \PhiT'} {\omega \kB^2 \kT'} \left(\kB^2 v_F^2 +\frac{1-v_F^2}{1-v^2} \vB^2 k_3^2 \right),\ann
	i_3 &= \frac{\eta_\gr \PhiT'}{\kB^2 \kT'}  \frac{1-v_F^2}{1-v^2} \left(\kB^2 - \omega(\kB\vB)\right),
\end{align}
and
\begin{equation*}
	\PhiT' = \PhiT\left(\frac{\kT'}{2m}\right),\ k_3^2 = \omega^2 - \kB^2.  
\end{equation*}
For $v=0$ we obtain conductivity of graphene at rest \eqref{eq:eta1}. 

\section{The Casimir energy and Force}\label{Sec:Casimir}

In the frameworks of the scattering matrix approach \cite{Fialkovsky:2018:qfcrbcs} the Casimir energy  per unit area for real frequencies reads 
\begin{equation}\label{eq:Efirst}
	\mathcal{E}  = - \frac{1}{2 \ii } \iint \frac{\dd^2 k}{(2\pi)^3}\left( I_- - I_+\right),
\end{equation}
where 
\begin{equation}
	I_\pm = \int_0^\infty \dd \omega \ln \det  \left[ 1 - e^{ \pm 2 \ii a k_3} \bm{\RC}(\pm k_3) \right],
\end{equation}
$\bm{\RC}(\pm k_3) = \bm{r}'_1(\pm k_3) \bm{r}_2(\pm k_3)$, and $k_3 = \sqrt{\omega^2 - \kB^2}$. The Eq. \eqref{eq:Efirst} contains all contributions to the Casimir energy -- due to evanescent and waveguide surface contributions ($\omega <k$) and propagating waves ($\omega >k$). The subscript $1(2)$ means that all reflection matrices related to the left (right) subsystem ($z=0,a$). Two terms $I_\pm$ correspond to opposite directions of scattering. 

The scattering matrix for each part of the system has the following form 
\begin{equation*}
	\bm{\mathcal{S}} = 
	\begin{pmatrix}
		\bm{r} & \bm{t}' \\
		\bm{t} & \bm{r}'
	\end{pmatrix},
\end{equation*}
with the corresponding index and argument.  This matrix for the conductive plane at position $z=a$ was found in Ref.\,\cite{Fialkovsky:2018:qfcrbcs}
\begin{equation}
	\bm{r} = - e^{2\ii k_3 a} \bm{r}_0, \bm{r}' = - e^{-2\ii k_3 a} \bm{r}_0, \bm{t} = \bm{t}' = \IB + \bm{r}_0,\label{eq:r}
\end{equation}
where 
\begin{equation*} 
	\bm{r}_0 = \frac{\omega^{2} \etaB - \kB \otimes (\kB\etaB) + \IB \omega k_3 \det\etaB}{\omega^2 \tr \etaB - \kB\kB\etaB + \omega k_3 (1 + \det \etaB)},
\end{equation*}
is the reflection matrix for plane at position $z =0$. 

The pressure is the negative derivative of the energy per unit area with respect to distance $\mathcal{P} = - \mathcal{E}'_a$. With simple algebra we obtain
\begin{equation}\label{eq:Ffirst}
	\mathcal{P}  = \iint \frac{\dd^2 k}{(2\pi)^3}\left( J_- +  J_+\right),
\end{equation}
where 
\begin{equation}
	J_\pm = \int_0^\infty \dd \omega k_3 \frac{e^{ \pm 2 \ii a k_3} (\tr \bm{\RC}(\pm k_3)  - 2e^{ \pm 2\ii a k_3} \det \bm{\RC}(\pm k_3))}{\det  \left[ \IB - e^{ \pm 2 \ii a k_3} \bm{\RC}(\pm k_3) \right]}.
\end{equation}

\begin{figure}[ht]\centering
	\includegraphics[width = 5cm]{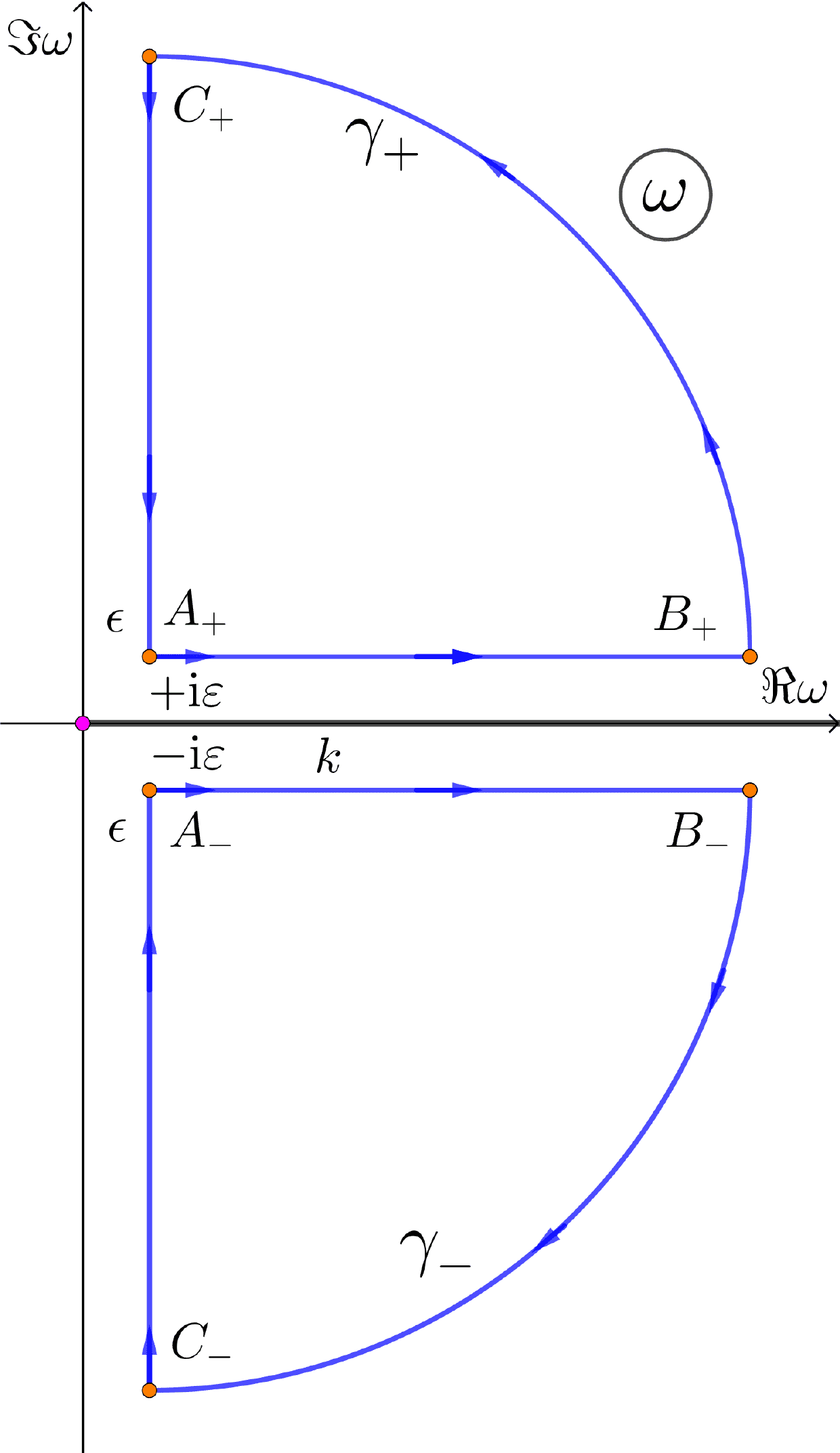}
	\caption{Contours of integrations.}\label{fig:path}
\end{figure}

The energy \eqref{eq:Efirst} and the pressure \eqref{eq:Ffirst} have real and imaginary contributions due to velocity. The Casimir friction has to appear as an imaginary part of the energy. The Casimir energy for real frequencies has been considered in Ref. \cite{Henkel:2004:cspatcf}. To separate real and imaginary parts we shift the integration from the real axis to the imaginary one as shown in Fig.\,\ref{fig:path}. The sign of phase of $k_3$ in the complex plain of $\omega$ coincides with the sign of phase of $\omega$ and defines the regions where the exponents become damping. 
Therefore, 
\begin{equation}
	I_\pm :  \int_0^\infty \Rightarrow \oint_{\gamma_\pm}  - \int_{C_\pm A_\pm}  - \int_{B_\pm C_\pm} \Rightarrow \sum_{i=1}^3 \mathcal{E}_i,
\end{equation}
where limits, $\epsilon,\varepsilon \to 0$, and radius of arcs, $R\to \infty$, are assumed.  The contributions of arcs $B_\pm C_\pm$ are zeros in the last limit, $\mathcal{E}_3 \to 0$. Therefore, the energy is a sum of two contributions $\mathcal{E} = \mathcal{E} _1 + \mathcal{E}_2$. Let us consider consecutively these terms separately. 

\subsection{$\mathcal{E}_2$: The real contribution}\label{Sec:Casimir2}

The second integral alongside the imaginary axis reads 
\begin{align}
	\mathcal{E}_2=& \frac{1}{2 \ii } \iint \frac{\dd^2 k}{(2\pi)^3} \left\{ \int_{C_-A_-} \dd \omega \ln \det  \left[ 1 - e^{ - 2 \ii a k_3} \bm{\RC}(- k_3) \right] \right. \ann
	-& \left.\int_{C_+A_+} \dd \omega \ln \det  \left[ 1 - e^{ 2 \ii a k_3} \bm{\RC}(k_3) \right] \right\}. 
\end{align}
One changes integrand variable $\omega=\ii \xi + \epsilon$. In this case $k_3 = \ii k_E\sign (\xi) $, where $k_E = \sqrt{\xi^2 + k^2}$. Therefore, 
\begin{align}
	\mathcal{E}_2=& \frac{1}{2} \iint \frac{\dd^2 k}{(2\pi)^3} \left\{ \int_{-\infty}^0 \dd \xi \ln \det  \left[ 1 - e^{ - 2 a k_E} \bm{\RC}(\ii k_E) \right] \right. \ann
	-& \left.\int^0_{+\infty} \dd \xi \ln \det  \left[ 1 - e^{ - 2  a k_E} \bm{\RC}(\ii k_E) \right] \right\}, 
\end{align}
and we arrive with formula \cite{Fialkovsky:2018:qfcrbcs}
\begin{equation}
	\mathcal{E}_2  =  \iint \frac{\dd^2 k}{2(2\pi)^3} \int_{-\infty}^{+\infty} \dd\xi  \ln \det  \left[ 1 - e^{ -2 a k_E}  \bm{\RC}(\ii k_E) \right]. 
\end{equation}

Appendix \ref{Sec:anprop} shows that the conductivities of graphene obey the general relation \cite[\S 82]{Landau:1984:EoCM} which must be  fulfilled for any realistic conductivity 
\begin{align}
	\etaB_1 (-\omega^*) &= \etaB_1^* (\omega), \ann
	\etaB_2 (-\omega^*,\vB) &= \etaB^*_2(\omega,-\vB).
\end{align}
Taking into account these relations, only, it is easy to show that the $\mathcal{E}_2$ is real (see Appendix \ref{Sec:anprop}) and
\begin{equation}
	\mathcal{E}_2  =  \Re \iint \frac{\dd^2 k}{(2\pi)^3} \int_0^{\infty} \dd\xi   \ln \det  \left[ \IB - e^{ -2 a k_E}  \bm{\RC}(\ii k_E) \right]. \label{eq:E2}
\end{equation}
This expression has represented in Ref. \cite{Fialkovsky:2018:qfcrbcs} in the following manifest form
\begin{align}
	\mathcal{E}_2 &= \int \frac{\dd^2 k}{(2\pi)^3} \int_0^\infty \dd\xi \ln \left(1 + e^{-4 a k_E} \frac{\xi^2 k_E^2 }{b_1b_2} \det \etaB_1 \det \etaB_2 \right. \ann
&-\left.  e^{-2 a k_E}\left[\frac{\xi^2 k_E^2 }{b_1b_2} \left[(1- \det\etaB_1) (1- \det\etaB_2) + \det(\etaB_1 - \etaB_2) \right] \right.\right.\ann 
&\left.\left.- \frac{\xi k_E}{b_1}   - \frac{\xi k_E}{b_2}   + 1 \right] \right),\label{eq:XY}
\end{align}
where $b_i = \xi^2 \tr \etaB_i + (\kB\kB\etaB_i)  + \xi k_E \bigl(1+ \det\etaB_i\bigr)$.

\subsection{$\mathcal{E}_1$: The Casimir friction}\label{Sec:Casimir1}

Let us proceed to the last contribution which comes from the integrals over closed contours $\gamma_\pm$
\begin{align}
	\mathcal{E}_1=& - \frac{1}{2 \ii } \iint \frac{\dd^2 k}{(2\pi)^3} \left\{  \oint_{\gamma_-} \dd \omega \ln \det  \left[ \IB - e^{ - 2 \ii a k_3} \bm{\RC}(-k_3) \right]  \right.\ann
	- &\left. \oint_{\gamma_+} \dd \omega \ln \det  \left[ \IB - e^{ 2 \ii a k_3} \bm{\RC}(k_3) \right] \right\}. 
\end{align}
This part is the source of the Casimir friction because it is imaginary.  The contribution to pressure reads
\begin{align}
	&\mathcal{P}_1  =\iint \frac{\dd^2 k}{(2\pi)^3}\ann
	\times & \left\{  \oint_{\gamma_-} \dd \omega k_3 \frac{e^{ - 2 \ii a k_3} (\tr \bm{\RC}(- k_3)  - 2e^{ - 2\ii a k_3} \det \bm{\RC}(- k_3))}{\det  \left[ \IB - e^{ - 2 \ii a k_3}\bm{\RC}(- k_3) \right]}\right.\ann
	+ &\left. \oint_{\gamma_+} \dd \omega k_3 \frac{e^{2 \ii a k_3} (\tr \bm{\RC}(k_3) - 2e^{2 \ii a k_3} \det \bm{\RC}(k_3))}{\det  \left[ \IB - e^{2 \ii a k_3} \bm{\RC}(k_3) \right]}\right\}. 
\end{align}
Therefore, the imaginary part of force may appear due to zeros of the denominator 
\begin{equation}\label{eq:poles}
	\det  \left[ \IB - e^{ \pm 2 \ii a k_3} \bm{r}'_1(\pm k_3) \bm{r}_2(\pm k_3) \right] = 0,
\end{equation}
as residues in these points. 

For the scalar Fresnel coefficients these poles have been noted by Lifshitz in Ref.\,\cite{Lifshitz:1956:tmafbs} when the integration over real axes is changed to the imaginary one (see also \cite{Kampen:1968:otmtovdwf,*Gerlach:1971:eovdwfbsatspi}). In the Ref.\,\cite{Henkel:2004:cspatcf} (see also review \cite{Joulain:2005:sewterhtcpacfritnf}) noted that the roots of this relation in the scalar case may appear in the complex plane, only. Lifshitz \cite{Lifshitz:1956:tmafbs} proved an absence of poles for normal dielectrics.  

Let us suppose that the matrix $\bm{\RC}(\pm k_3) = \bm{r}'_1(\pm k_3) \bm{r}_2(\pm k_3)$ has eigenvalues $  r_\tm^2(\pm k_3)$ and $r_\te^2(\pm k_3)$, which means that the following relation is valid
\begin{equation}\label{eq:eigen}
	\bm{\RC}(\pm k_3)= \TB^{-1} \DB_\pm \TB, 
\end{equation}
where
\begin{equation}
	\DB_\pm = 
	\begin{pmatrix}
		r_\tm^2 (\pm k_3)& 0 \\
		0& r_\te^2 (\pm k_3)
	\end{pmatrix}.
\end{equation}

For $v=0$, the $r_\tm^2$ and $r_\te^2$ are the squares of reflection coefficients of TM and TE modes respectively for a system of two planes with the corresponding sign of $k_3$.  

Taking into account Eq.\,\eqref{eq:eigen} and relation 
\begin{equation*}
	\oint_{\gamma_-} \dd \omega k_3 + \oint_{\gamma_+} \dd \omega  k_3 = 0,
\end{equation*}
we obtain
\begin{align}
	&\mathcal{P}_1= \iint \frac{\dd^2 k}{(2\pi)^3} \ann 
	&\left\{  \oint_{\gamma_-} \dd \omega k_3\left[ \frac{1 }{1 - e^{ - 2 \ii a k_3} r_\tm^2(-k_3)} + \frac{1}{1- e^{ - 2 \ii a k_3} r_\te^2(-k_3)}\right] \right.\ann
	+ &\left. \oint_{\gamma_+} \dd \omega  k_3 \left[ \frac{1}{1- e^{ 2 \ii a k_3} r_\tm^2(k_3)} + \frac{1}{1- e^{ 2 \ii a k_3} r_\te^2(k_3)}\right] \right\}. 
\end{align}

Therefore,  the contributions to these integrals come from the roots 
\begin{equation}\label{eq:zeros}
	1 - e^{ \pm 2 \ii a k_3} r_\tm^2(\pm k_3) = 0,\ 1 - e^{ \pm 2 \ii a k_3} r_\te^2(\pm k_3) = 0,
\end{equation}
as residues in these poles if the poles are inside the corresponding contour. For real frequencies, these relations describe the waveguide modes \cite{Bordag:2012:evetpstswapm}, which are the evanescent modes with $\omega <k$.  A similar relation was considered in Ref.\,\cite{Silveirinha:2014:oiaslepmm}. 

Let us denote the solutions of equations 
\begin{equation}
	1 - e^{ -2 \ii a k_3}r_\tm^2(-k_3) = 0,\ 1 - e^{ - 2 \ii a k_3} r_\te^2(-k_3)= 0,
\end{equation}
by $\omega^{n-}_\tm, \omega^{n-}_\te$ with negative imaginary parts $\Im\omega^{n-}_\tm <0$, $\Im \omega^{n-}_\te<0$, and the solutions of equations 
\begin{equation}
	1 - e^{ 2 \ii a k_3} r_\tm^2(k_3)= 0,\ 1 - e^{ 2 \ii a k_3} r_\tm^2 (k_3)= 0,
\end{equation}
by $\omega^{n+}_\tm, \omega^{n+}_\te$ with positive imaginary parts $\Im\omega^{n+}_\tm >0$, $\Im \omega^{n+}_\te>0$. Then, the pressure reads 
\begin{align}
	\mathcal{P}_1&= \ii \sum_n\iint \frac{\dd^2 k}{(2\pi)^2} \ann
	\times& \left[\us{\Res}{\omega=\omega^{n+}_\tm} \frac{k_3}{1- e^{ 2 \ii a k_3} r_\tm^2(k_3) } - \us{\Res}{\omega=\omega^{n-}_\tm} \frac{k_3}{1 - e^{ - 2 \ii a k_3} r_\tm^2(-k_3)}\right.\ann
	+&\left. \us{\Res}{\omega=\omega^{n+}_\te} \frac{k_3}{1- e^{ 2 \ii a k_3} r_\te^2(k_3)}  - \us{\Res}{\omega=\omega^{n-}_\te}\frac{k_3 }{1- e^{ - 2 \ii a k_3} r_\te^2(-k_3)} \right].
\end{align}

Let us analyze the general properties of existing poles in the contours following by Lifshitz \cite{Lifshitz:1956:tmafbs}. We represent the relations \eqref{eq:zeros} in the following form 
\begin{equation}\label{eq:zeros1}
	e^{ \pm 2 \ii a k_3} = r_\tm^{-2}(\pm k_3),\ e^{ \pm 2 \ii a k_3} = r_\te^{-2}(\pm k_3),
\end{equation}
and consider the modulo of both sides. 

It is easy to see that $\sign (\Im k_3) = \sign (\Im \omega)$. Therefore, the modulo of the left-hand side of Eqs. \eqref{eq:zeros1} reads
\begin{equation}
	\left|e^{ \pm 2 \ii a k_3}\right| = \left(e^{ \pm 2 \ii a k_3} e^{ \mp 2 \ii a k_3^*}\right)^{1/2} =  e^{\mp 2a \Im k_3} = e^{-2a |\Im k_3|} <1. 
\end{equation}
Therefore, the solutions exist if and only if the relations
\begin{equation}\label{eq:poles1}
	\left|r_\tm^{-2}(\pm k_3)\right| <1, \left|r_\te^{-2}(\pm k_3)\right| <1,
\end{equation}
are valid, or 
\begin{equation}\label{eq:poles2}
	\left|r_\tm^2(\pm k_3)\right| >1, \left|r_\te^2(\pm k_3)\right| >1.
\end{equation}
It means that it has to be photon production. 

Let us consider now the threshold of velocity for the Casimir friction. It was found in the Ref.\,\cite{Farias:2017:QFbGS,*Farias:2015:fatqfeaadf} in the framework of effective action. Here we prove this threshold by using the simple observation that the origin for the Casimir friction is the mixing between positive and negative frequencies \cite{Maghrebi:2013:qcranf}.

For graphene at rest, $v=0$, the spectrum is defined by relation $\kT= \sqrt{\omega^2 - v_F^2 k^2} = 0$. We obtain two branches 
\begin{equation}
	\omega^\pm_0/k = \pm v_F. 
\end{equation}

For moving graphene the spectrum is defined by relation $\kT' = \sqrt{\omega'^2 - v_F^2 k'^2}  = 0$. We have a quadratic equation
\begin{equation}
	\omega^2 - v_F^2 k^2 + \frac{1-v_F^2}{1-v^2} (\omega^2 v^2 + (\kB\vB)^2 -2 \omega (\kB\vB) ) =0,
\end{equation}
with solutions 
\begin{align}
	\omega^\pm_v/k &=  v\frac{1-v_F^2}{1-v^2 v_F^2}\cos \varphi   \ann 
	\pm& v_F \frac{\sqrt{1-v^2}}{1-v^2 v_F^2} \sqrt{1- v^2 v_F^2 - v^2 (1-v_F^2) \cos ^2\varphi}.\label{eq:spectrum}
\end{align}
The expression under the square root is always positive.  Here, the $\varphi$ is the angle between vectors $\kB$ and $\vB$. It is easy to verify that $\lim_{v\to 0} \omega_v^\pm = \omega^\pm_0$.  

To find  a condition for mixing the frequencies let us first for simplicity expand \eqref{eq:spectrum} over velocity $v <1$. We set $v = v_F \beta$ and expand over $v_F \ll 1$:
\begin{equation}
	\omega_v^\pm/k = v_F (\pm 1 +\beta  \cos \varphi) + O(v_F^3).
\end{equation}
For $\beta <1\ (v<v_F)$ we observe that $\omega_v^+ >0$ and $\omega_v^- <0$ for all $\varphi$, that is there is no mixing between modes. In the opposite case $\beta >1\ (v>v_F)$, there are domains of angle $\varphi$ where  $\omega_v^+ <0$ or $\omega_v^- >0$ -- the mixing appears and therefore the Casimir friction appears.  

This statement is valid without expansion over $v$. Indeed, the condition $\omega_v^+<0$ reads 
\begin{equation*}
	v_F \frac{\sqrt{1-v^2}}{1-v^2 v_F^2} \sqrt{1- v^2 v_F^2 - v^2 (1-v_F^2) \cos ^2\varphi} <-v\frac{1-v_F^2}{1-v^2 v_F^2}\cos \varphi.
\end{equation*}
The first observation is that $\varphi \in \left[\frac{\pi}{2}, \frac{3\pi}{2} \right]$, because expression on right hand side should be positive.  With this limitation, one can put in a square on both sides and one obtains a condition 
\begin{equation}
	\cos ^2\varphi > 1-\frac{v^2-v_F^2}{v^2 \left(1- v_F^2\right)}.
\end{equation}
We observe that this relation can not be valid if $v<v_F$. For $v>v_F$ there are domains of $\varphi$ where solutions exist and therefore the mixing of modes exists, and the Casimir friction comes up. 

The condition $\omega_v^->0$ reads 
\begin{equation*}
	v_F \frac{\sqrt{1-v^2}}{1-v^2 v_F^2} \sqrt{1- v^2 v_F^2 - v^2 (1-v_F^2) \cos ^2\varphi} <v\frac{1-v_F^2}{1-v^2 v_F^2}\cos \varphi.
\end{equation*}
Therefore, $\varphi \in \left[-\frac{\pi}{2},\frac{\pi}{2}\right]$.  With this limitation one can put in a square both sides and one obtains the same condition 
\begin{equation}
	\cos ^2\varphi > 1-\frac{v^2-v_F^2}{v^2 \left(1- v_F^2\right)},
\end{equation}
which has solutions for $v > v_F$, only.  
\begin{figure}[h]
	\centering
	\includegraphics[width=0.5\linewidth]{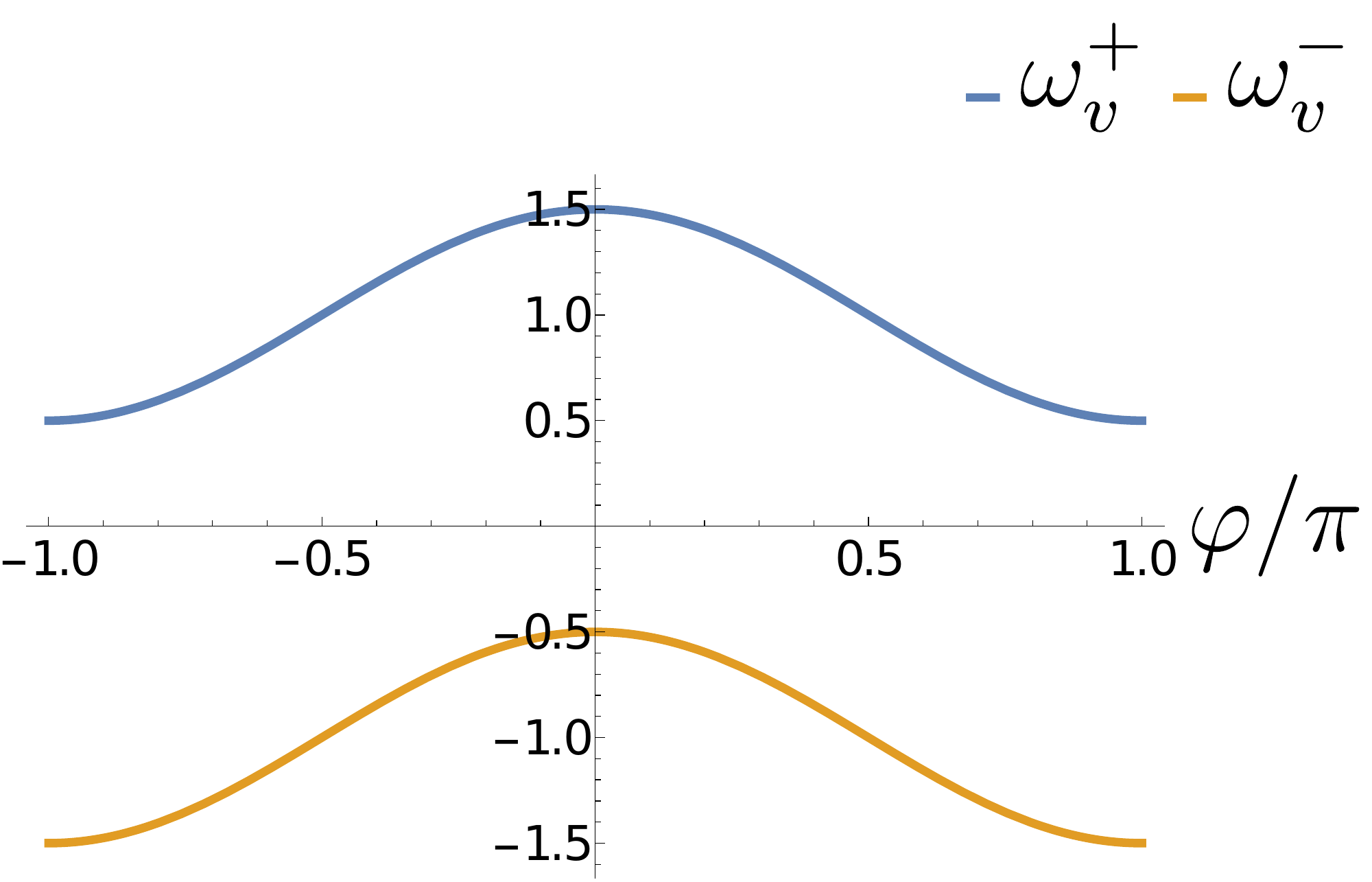}\includegraphics[width=0.5\linewidth]{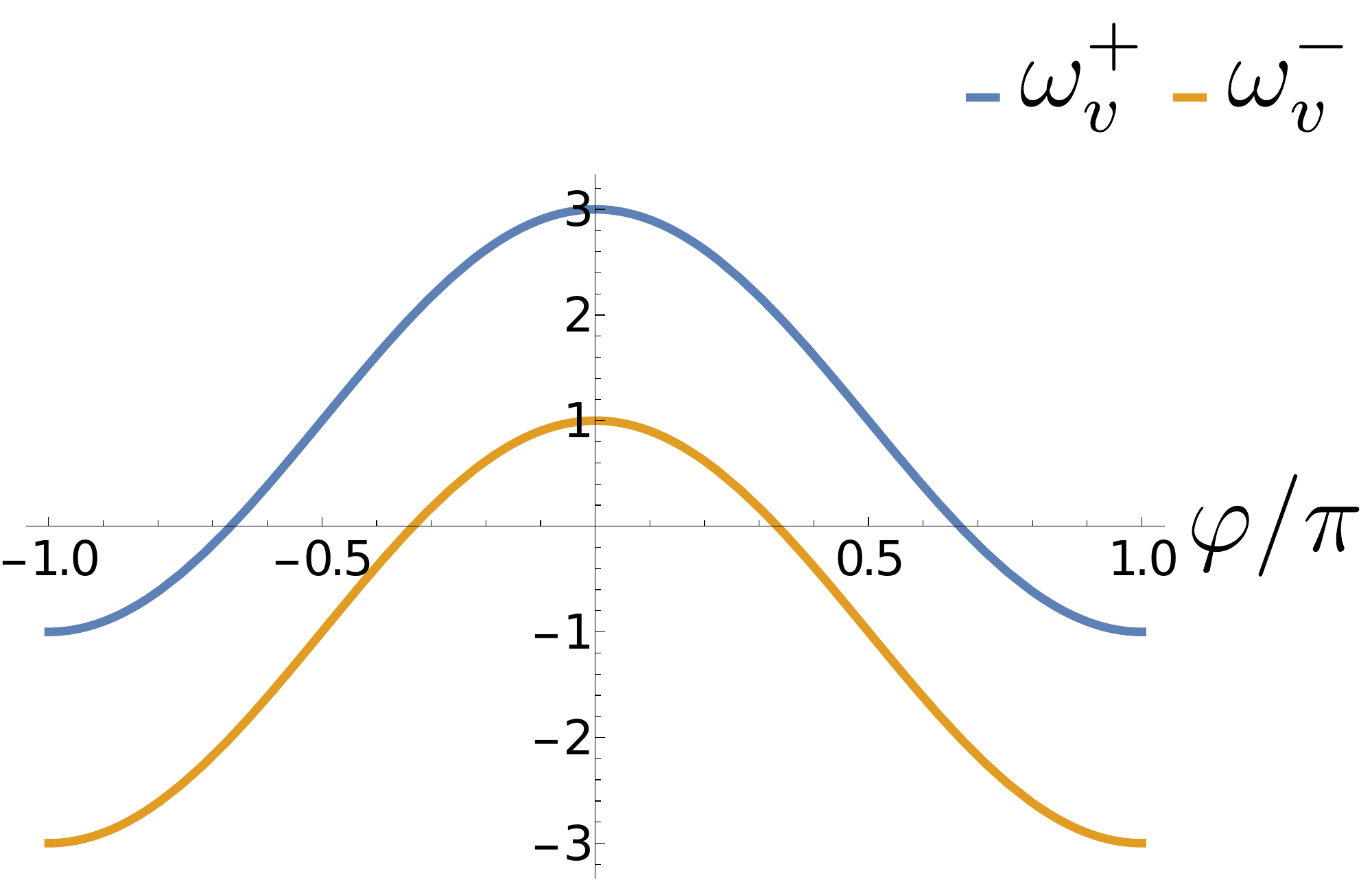}
	\caption{The spectrum \eqref{eq:spectrum}. Left panel: $v/v_F = 0.5$ (there is no mixing). Right panel: $v/v_F=2$ (there is mixing).}
	\label{fig:mix}
\end{figure}

Let us consider threshold for photon production. It happens if \cite{Maghrebi:2013:qcranf,*Maghrebi:2013:sadce}
\begin{equation}
	\omega^+_v + \omega^+_0 <0.
\end{equation}
From Fig.\,\ref{fig:mix} we observe that we have to consider this relation for $\varphi = \pm \pi$. It gives threshold for velocity
\begin{equation}
	v > \frac{2v_F}{1+v_F^2} \approx 2v_F,
\end{equation}
which is close to that observed in Refs.\,\cite{Maghrebi:2013:qcranf,*Maghrebi:2013:sadce}. 

The detailed analysis of the Casimir friction we will make elsewhere in a separate paper. 

\section{The Casimir energy: velocity correction} \label{Sec:Real}

Let us consider the contribution $\mathcal{E}_2$ to the Casimir energy given by Eq.\,\eqref{eq:E2} for two systems -- graphene/graphene (gg) and graphene/ideal metal (gm). We consider first the zero-velocity energy $\mathcal{E}_0 \equiv \mathcal{E}_2|_{v=0}$ and then the velocity correction which we will denote $\Delta_v \mathcal{E}_2 = \mathcal{E}_2 - \mathcal{E}_0$. Also, we will give expressions for velocity correction to the Casimir pressure $\Delta_v \mathcal{P}_2 = \mathcal{P}_2 - \mathcal{P}_0$.

\subsection{Two graphene}

The zero-velocity Casimir energy reads
\begin{align}
	\mathcal{E}_0^{\gs\gs} &= \int_0^\infty \frac{k_E^2 \dd k_E}{(2\pi)^2} \int_0^1 \dd x \ann 
	&\times \left\{\ln \left(1 - e^{-2a k_E} r_\tm^2\right)  + \ln \left(1 - e^{-2a k_E} r_\te^2\right)\right\}, \label{eq:E0gg}
\end{align}
where 
\begin{equation}
	r_\tm^{-1} = 1 + \frac{t}{\eta_\gr \Phi\left(\frac{k_E t}{2m a}\right)},\ r_\te^{-1} = -1 - \frac{1}{ \eta_\gr t \Phi\left(\frac{k_E t}{2m a}\right)},
\end{equation}
are the reflection coefficients for TM and TE modes \cite{Bordag:2009:CibapcagdbtDm}.  Here, $\Phi(y) = \PhiT (\ii y)$, and we used spherical coordinates, $k_E = \sqrt{\xi^2 + k^2}$, $x = \cos\theta$, and $t = \sqrt{x^2 + v_F^2 (1-x^2)}$.  Let us estimate the $\mathcal{E}_0^{\gs\gs} $ for large and small separations. 

For $a\to\infty$, the function $\Phi (k_E t/2ma) \approx 4 k_E t/3\pi ma$ and we obtain
\begin{align}
	\frac{	\mathcal{E}_0^{\gs\gs} }{	\mathcal{E}_{\id} } &\approx \frac{32 \eta_\gr^2 \left(9 + 2 v_F^2 + 4 v_F^4\right)}{\pi ^6 (am)^2} \approx \frac{288 \eta_\gr^2}{\pi ^6 (am)^2},\ann
	\frac{	\mathcal{P}_0^{\gs\gs} }{\mathcal{P}_{\id} } &\approx \frac{160 \eta_gr^2 \left(9 + 2 v_F^2+4 v_F^4\right)}{3 \pi ^6 (am)^2} \approx \frac{480 \eta_\gr^2}{\pi ^6 (am)^2},
\end{align}
where $\mathcal{E}_{\id} = - \pi^2/720 a^3$ is the Casimir energy for two ideal plates. Therefore, for large separation $\mathcal{E}_0^{\gs\gs} \approx - 2\eta_\gr^2/5\pi^4 a^5m^2$. 

For $a\to 0$, the function $\Phi \to 1$ (all expressions see in Appendix \ref{Sec:App2}).  Numerically, $\mathcal{E}_0^{\gs\gs}/	\mathcal{E}_{\id} \approx 0.0048$, for $v_F = 1/300$, and therefore $\mathcal{E}_0^{\gs\gs} \sim 1/a^3$. 

Note, that for constant conductivity case \cite{Khusnutdinov:2014:Cefswcc} the energy is proportional to the first degree of $\eta_\gr$ due to TM mode and falls as $1/a^3$ for any separations. The constant conductivity case is realized for $v_F=0$ and $\Phi =1$ (see Eq.\,\eqref{eq:eta1}). But setting $v_F=0$ does not lead to relation $\Phi =1$. The dispersion over angle $\theta$ is still survived. The case $\Phi =1$ is realized for small separation. By setting $v_F =0$ in this relation we obtain $\mathcal{E}_0^{\gs\gs}/\mathcal{E}_{\id}  \to 45 \eta_{\gr}/\pi^4$ which coincides with result of Ref.  \cite{Khusnutdinov:2014:Cefswcc} for constant conductivity case. The large separation case is completely different from the constant conductivity case -- the dispersion over angle $\theta$ and non-zero gap play an important role. 

The velocity correction term $\mathcal{E}_2^{\gs\gs}$ has a complicated form and no reason to reproduce it here. For large separation, $(am)\to \infty$ and arbitrary velocity
\begin{align}\label{eq:Large}
	\frac{\Delta_v\mathcal{E}_2^{\gs\gs}}{\mathcal{E}_{\id}} &\approx  \frac{112 \eta_\gr^2 (1-v_F^2)^2}{\pi^6 (am)^2 } \frac{v^2}{1-v^2} \approx  \frac{112 \eta_\gr^2}{\pi^6 (am)^2 }\frac{ v^2}{1-v^2},\ann
	\frac{\Delta_v\mathcal{P}_2^{\gs\gs}}{\mathcal{P}_{\id}} &\approx  \frac{560 \eta_\gr^2 (1-v_F^2)^2}{3\pi^6 (am)^2 } \frac{v^2}{1-v^2} \approx  \frac{560 \eta_\gr^2}{3\pi^6 (am)^2 }\frac{ v^2}{1-v^2}.
\end{align}
and for small separation, $(am) \to 0$ and small velocity $v\ll v_F$
\begin{equation}\label{eq:Small}
	\frac{\Delta_v\mathcal{E}_2^{\gs\gs}}{\mathcal{E}_{\id}} \approx C_e v^2,\ \frac{\Delta_v\mathcal{P}_2^{\gs\gs}}{\mathcal{P}_{\id}} \approx C_p v^2.  
\end{equation}
For $v_F = 1/300$ one has $C_e= C_p = 1.396$. Therefore, $\mathcal{E}_2^{\gs\gs} \sim a^{-5}$ for large separations and $\mathcal{E}_2^{\gs\gs} \sim a^{-3}$ for small ones.  In the non-relativistic case $v \ll v_F$ one has  
\begin{equation}
	\frac{\Delta_v\mathcal{E}_2^{\gs\gs}}{\mathcal{E}_0^{\gs\gs}} = E(am) v^2,\ \frac{\Delta_v\mathcal{P}_2^{\gs\gs}}{\mathcal{P}_0^{\gs\gs}} = P(am) v^2. 
	\label{eq:Asym}
\end{equation}
Due to Eqs.\,\eqref{eq:Large}, \eqref{eq:Small}, the function $E(ma)$ has the following limits 
\begin{equation}\label{eq:Asym1}
	E(ma) = 
	\begin{cases}
		\frac{7}{18},\ ma\to \infty\\
		290, \ ma\to 0
	\end{cases}. 
\end{equation}
The function $P$ has the same values for the limit cases. 

\begin{figure}
	\centering
	\includegraphics[width=1\linewidth]{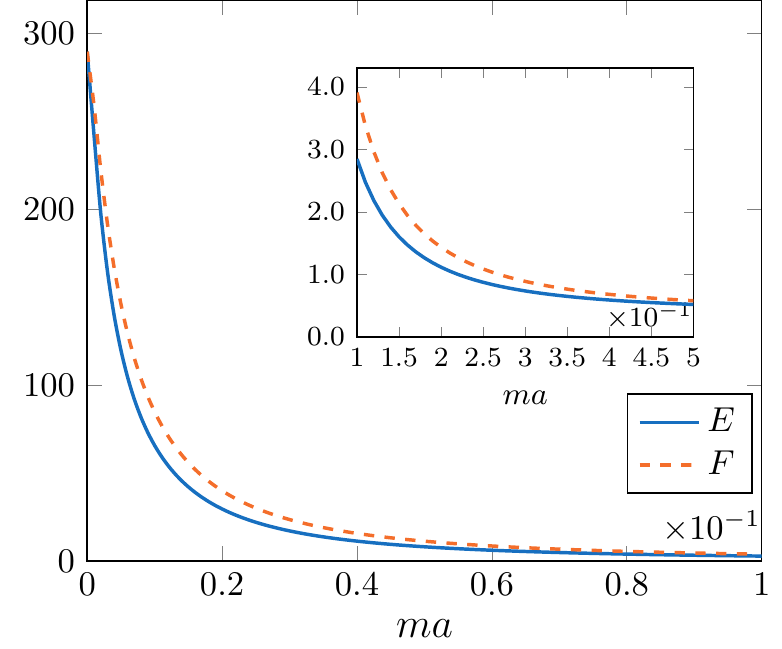}
	\caption{The functions $E$ and $F$ \eqref{eq:Asym} describe the velocity correction. Here we used $v_F = 1/300$. The limits $a\to 0$ and $a\to\infty$ are in accordance with Eq.\,\eqref{eq:Asym1}.}
	\label{fig:f2}
\end{figure}

Numerical evaluations of $E(ma)$ are shown in Fig.\,\ref{fig:f2}.
The contribution $\mathcal{E}_2$ to the Casimir energy $\mathcal{E}$ is suppressed by factor $(\vB/c)^2$. 

\begin{figure}
	\centering
	\includegraphics[width=0.5\linewidth]{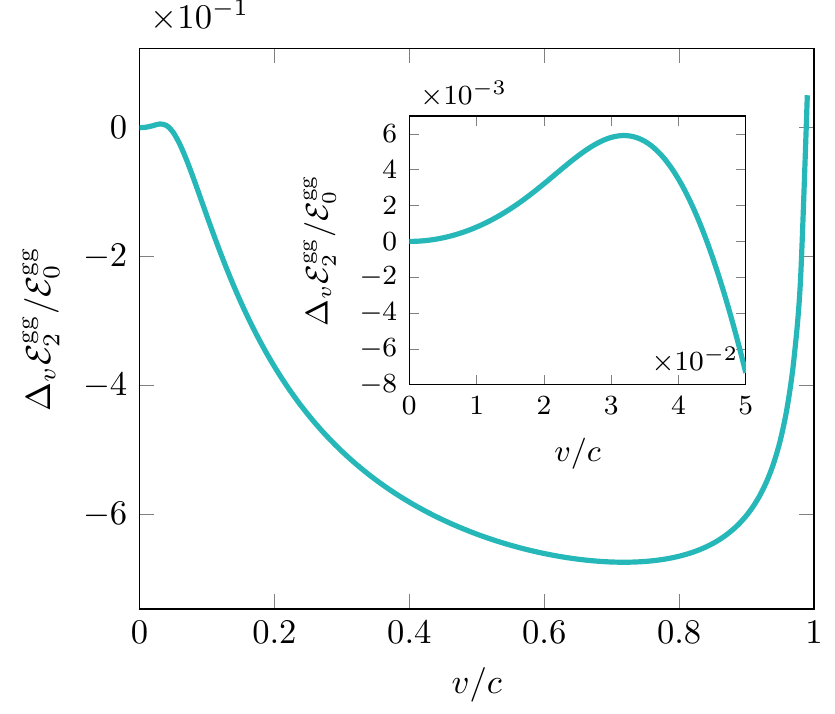}\includegraphics[width=0.5\linewidth]{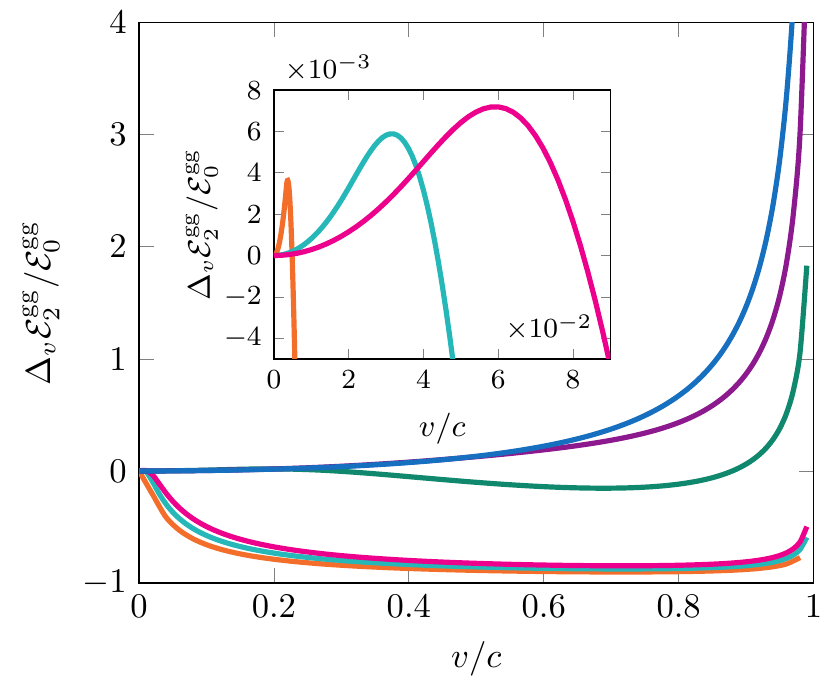}
	\caption{The numerical evaluation the relative correction $\Delta_v \mathcal{E}_2^{\gs\m} /\mathcal{E}_0^{\gs\m}= (\mathcal{E}_2^{\gs\m} - \mathcal{E}_0^{\gs\m})/\mathcal{E}_0^{\gs\m}$ for Casimir energy \eqref{eq:E2} for case of two graphene. Here we used $v_F = 1/300$ and $\eta_\gr = 0.0114$. Left panel: $ma = 0.05$ ($m=0.1$ (eV), $a=100$ (nm)). The quadratic behavior \eqref{eq:Asym} occurs for small velocities $v \ll v_F$. Right panel: $ma = 0,0.05,0.1,0.3,0.9,2$ (from the bottom upwards). The maxima occurred for critical velocity \eqref{eq:vcr}.}
	\label{fig:gg}
\end{figure}

The numerical evaluation of exact expression \eqref{eq:E2} is shown in Fig.\,\eqref{fig:gg}.  
We observe dependence $\sim v^2$ for domain $v \ll v_F$. In the ultrarelativistic case, the energy is divergent.  The energy has a maximum at specific velocity $v_{\rm cr}^{\gs\gs}$. The numerical evaluations reveal the following dependence 
\begin{equation}\label{eq:vcr}
	v_{\rm cr}^{\gs\gs} - v_F\approx  \frac{1}{2} (ma), 
\end{equation}  
for $ma <1$. The velocity Fermi defines the position of the $v_{\rm cr}^{\gs\gs}$. In the gapeless case $v_{\rm cr}^{\gs\gs} = v_F$. 

\subsection{Graphene and ideal metal}

In this case, we have to take limit $\eta_\gr \to \infty$ for graphene at rest in Eq. \eqref{eq:E0gg}. In this limit $r_\tm = 1$ and $r_\te = -1$.  Therefore, the zero-velocity term reads \cite{Bordag:2009:CibapcagdbtDm}
\begin{align}
	\mathcal{E}_0^{\gs \m} &= \int_0^\infty \frac{\kappa^2 \dd \kappa}{(2\pi)^2} \int_0^1 \dd x \ann 
	&\times \left\{\ln \left(1 - e^{-2a \kappa} r_\tm\right)  + \ln \left(1 + e^{-2a \kappa} r_\te\right)\right\}.\label{eq:E0gm}
\end{align}

For $a\to\infty$, we obtain 
\begin{equation}
	\left.\frac{\mathcal{E}_0^{\gs \m} }{	\mathcal{E}_{\id} } \right|_{a\to \infty} \approx \frac{60 \eta_\gr \left(2 + v_F^2 \right)}{\pi ^5 (a m)} \approx \frac{120 \eta_\gr}{\pi ^5 (a m)}.
\end{equation}
Therefore, for large separation $\mathcal{E}_0^{\gs \m} \sim \eta_\gr/a^4m$ \cite{Bordag:2009:CibapcagdbtDm}.  

For $a\to 0$ (see Appendix \ref{Sec:App2}) one has numerically	$\mathcal{E}_0^{\gs \m}/	\mathcal{E}_{\id} = 0.026$ for $v_F = 1/300$ and thus, $\mathcal{E}_0^{\gs \m} \sim 1/a^3$.   

\begin{figure}
	\centering
	\includegraphics[width=0.5\linewidth]{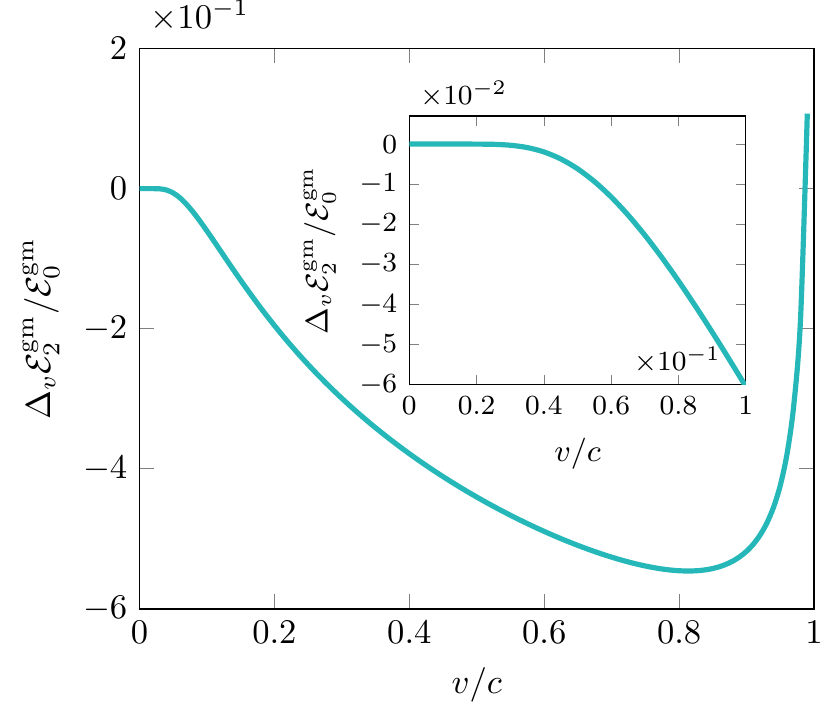}\includegraphics[width=0.5\linewidth]{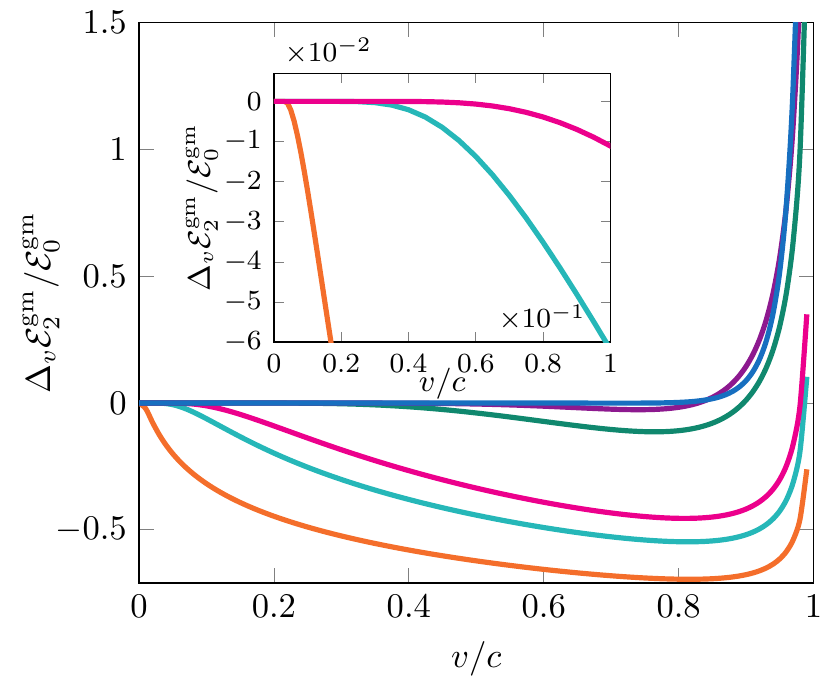}
	\caption{The numerical evaluation the relative correction $\Delta_v \mathcal{E}_2^{\gs\m}/\mathcal{E}_0^{\gs\m} = (\mathcal{E}_2^{\gs\m} - \mathcal{E}_0^{\gs\m})/\mathcal{E}_0^{\gs\m}$ for Casimir energy \eqref{eq:E2} for system ideal metal/graphene. Here we used $v_F = 1/300$ and $\eta_\gr = 0.0114$. Left panel: $ma = 0.05$ ($m=0.1$ (eV), $a=100$ (nm)). The correction is zero for $v < v_F$ in accordance with \eqref{eq:GMsmall}. Right panel: $ma = 0,0.05,0.1,0.5,0.9,2$ (from the bottom upwards).}
	\label{fig:gm}
\end{figure}

The numerical evaluations of  exact expression \eqref{eq:E2} are shown in Fig.\,\ref{fig:gm}.
We observe that for small velocities the energy is constant, and does not depend on the velocity up to some critical velocity $v_{\rm cr}^{\gs \m}$. Numerical analysis reveals the same linear dependence \eqref{eq:vcr} for critical velocity
\begin{equation}\label{eq:GMsmall}
	v_{\rm cr}^{\gs\m} - v_F\approx  \frac{1}{2} (ma),
\end{equation}  
for $ma<1$.

\section{Discussion and conclusion} 

We considered here the Casimir energy for two parallel graphenes with relative velocity $v$ in the framework of scattering theory. To obtain the conductivity tensor of moving graphene in the laboratory frame the Lorentz transformation of the boundary conditions is used, which defines the conductivity tensor in the terms of the polarization tensor. The Casimir energy consists of two parts -- real (Sec. \ref{Sec:Casimir2}) and imaginary  (Sec. \ref{Sec:Casimir1}). The real part is a  velocity correction to the usual Casimir energy and the imaginary one is the Casimir friction. For both of these energies, the Fermi velocity plays a crucial role. The Casimir friction is zero up to the Fermi velocity, for $v < v_F$. This fact is the simple consequence of the observation that the origin for the Casimir friction is the mixing between positive and negative frequencies \cite{Maghrebi:2013:qcranf} (see the end of Sec. \ref{Sec:Casimir1}) and was considered in Ref.\,\cite{Farias:2017:QFbGS,*Farias:2015:fatqfeaadf} in the framework of effective action approach. The Casimir friction appears if the Eqs.\,\eqref{eq:zeros} have corresponding solutions with a non-zero imaginary part of the frequency. It takes place if and only if the modulo of reflection coefficients are greater than one \eqref{eq:poles2}. The detailed evaluations of the Casimir friction will be considered in a separate paper. 

We considered here in detail the real part of the energy for two systems -- graphene/graphene and ideal metal/graphene with a relative velocity $v$.  In the first system, the velocity correction to the Casimir energy grows up as $v^2$ \eqref{eq:Asym} and then falls after some maximum value (\ref{fig:gg}). The position of this maximum is defined by the velocity Fermi and has linear dependence with $(ma)$ \eqref{eq:vcr}. For the system ideal metal/graphene, the energy correction is zero up to some value which has the same form \eqref{eq:GMsmall} and is defined by the velocity Fermi (see Fig.\,\ref{fig:gm}). 

For laboratory experiments, the velocity Fermi $v_F \approx 1/300 \approx 10^6$ m/s and situation $v \ll v_F$ takes place. The function $E$ in the Eq.\,\eqref{eq:Asym} has maximum value $290$ for gapless case, $m=0$. In this case and for $v=v_F$
\begin{equation}
	\frac{\Delta_v\mathcal{E}_2^{\gs\gs}}{\mathcal{E}_0^{\gs\gs}} = E(0) v_F^2 = 0.0033,
\end{equation} 
for system graphene/graphene and it is zero for the ideal metal/graphene configuration.  

\begin{acknowledgments}
NK was supported in part by the grants 2022/08771-5, 2021/10128-0 of S\~ao Paulo Research Foundation (FAPESP). One of us (NK) is grateful to M. Bordag, M. Silveirinha  and D. Vassilevich for fruitful discussions. MA acknowledges support from the French Agence Nationale pour la Recherche -- ANR ("CAT" project).
\end{acknowledgments}

\appendix 

\section{Analytical properties of conductivity} \label{Sec:anprop}

To calculate the Casimir energy we have to consider the conductivity as a function of complex frequency $\omega = \omega_r + \ii \omega_i$.  The general property of permittivity \cite[\S 82]{Landau:1984:EoCM} is
\begin{equation}
	\varepsilonB (-\omega^*) = \varepsilonB^*(\omega).
\end{equation}
For conductors $\varepsilonB(\omega) = \IB + 2 \ii \etaB (\omega) /\omega$, and  
\begin{equation}\label{eq:cond1}
	\etaB (-\omega^*) = \etaB^* (\omega).
\end{equation}

A graphene at rest has the following conductivity
\begin{equation}
	\etaB_1 = \eta_{\gr} \frac{\kT}{\omega} \left(\IB + v_F^2 \frac{\kB \otimes \kB}{\kT^2}\right) \PhiT \left(\frac{\kT}{2m}\right),
\end{equation}
where $\eta_\gr = 2\pi \sigma_\gr = \pi \alpha/2 = 0.0114$, and 
\begin{align*}
	\PhiT(y) &= \frac{2\ii }{\pi y}\left\{ 1 - \frac{y^2 + 1}{y}\arctanh y\right\},\\
	\kT &=  \sqrt{\omega^2 - v_F^2 \kB^2}. 
\end{align*}
Obviously, that $y(-\omega^*) = y(\omega^*) =y^*(\omega)$, and therefore 
\begin{equation}
	\arctanh y(-\omega^*) = \arctanh y^*(\omega) = \arctanh^* y(\omega),
\end{equation}
and
\begin{equation*}
	\PhiT(y(-\omega^*)) = - \PhiT^*(y(\omega)).
\end{equation*}
Therefore, the relation \eqref{eq:cond1} is valid: $\etaB_1 (-\omega^*) = \etaB_1^* (\omega)$. Then,   it is easy to verify that
\begin{equation}
	\etaB_1 (\omega^*) = -\etaB_1^* (\omega).
\end{equation}

At the imaginary axis $\omega = \ii \xi$ we obtain 
\begin{equation}\label{eq:cond3}
	\etaB_1 (\ii\xi) = \etaB_1^* (\ii \xi),\ \etaB_1 (-\ii\xi) = - \etaB_1^* (\ii \xi) = \etaB_1 (\ii\xi),
\end{equation}
and conductivity is real function 
\begin{equation}
	\etaB_1(\ii \xi) = \eta_{\gr} \frac{k_F}{\xi} \left(\IB - v_F^2 \frac{\kB \otimes \kB}{k_F^2}\right) \Phi \left(\frac{k_F}{2m}\right),
\end{equation}
where
\begin{align*}
	\Phi(y) &= \frac{2 }{\pi y}\left\{ 1 + \frac{y^2 - 1}{y}\arctan y\right\},\\
	k_F &=  \sqrt{\xi^2 + v_F^2 \kB^2}. 
\end{align*}
The conditions \eqref{eq:cond3} for function $\tilde{\etaB}_1(\xi) = \etaB_1(\ii \xi)$ read
\begin{equation}\label{eq:cond2}
	\etaBT_1 (\xi) = \etaBT_1 (\xi)^*,\ \etaBT_1 (-\xi) = - \etaBT_1 (\xi)^* = -\etaBT_1 (\xi).
\end{equation}

The conductivity  of the moving graphene has form \eqref{eq:eta2}
\begin{equation}
	\etaB _2 = \frac{\omega'}{\omega} \GB\etaB'_2 \GB^T,
\end{equation}
where 
\begin{equation}\label{eq:Gmat}
	\GB =\IB  - \frac{\omega}{(ku)} \frac{\uB\otimes \uB}{u_0+1} + \frac{\uB \otimes \kB}{(ku)}, 
\end{equation}
and
\begin{equation}
	\etaB'_2 = \eta_{\gr} \frac{\kT'}{\omega'} \left(\IB + v_F^2 \frac{\kB' \otimes \kB'}{\kT'^2}\right)\PhiT \left(\frac{\kT'}{2m}\right).
\end{equation}
Here, 
\begin{align}
	\kT' &= \sqrt{\omega'^2 - v_F^2 \kB'^2}\ann
	&= \sqrt{\kT^2 + \gamma^2 (1-v_F^2) (\omega^2 \vB^2 + (\kB\vB)^2 -2 \omega (\kB\vB) )},
\end{align}
and $\omega' = (ku) = \gamma (\omega - (\kB\vB))$. 

We observe now that 
\begin{equation*}
	\kT'(-\omega^*,\vB) = \kT'^*(\omega,-\vB),\ \omega'(-\omega^*,\vB) = -\omega'^*(\omega,-\vB),
\end{equation*}
and ($y' = \kT'/2m$)
\begin{equation*}
	\PhiT(y'(-\omega^*,\vB) ) = -\PhiT^*(y'(\omega,-\vB)).
\end{equation*}
Therefore, 
\begin{equation}
	\etaB'_2 (-\omega^*,\vB) = \etaB'^*_2(\omega,-\vB).
\end{equation}
It is easy to see from \eqref{eq:Gmat} that,
\begin{equation*}
	\GB(-\omega^*,\vB) = \GB'^*(\omega,-\vB)
\end{equation*} 
and therefore,  
\begin{equation}
	\etaB_2 (-\omega^*,\vB) = \etaB^*_2(\omega,-\vB).
\end{equation}
The conductivity of the moving graphene obeys the relation \eqref{eq:cond1} but with a corresponding changing direction of the velocity. It is similar to the case with an external magnetic field. 

Then, because 
\begin{align*}
	\kT'(\omega^*,\vB) &= \kT'^*(-\omega,-\vB) = \kT'^*(\omega,\vB),\ann
	\omega'(\omega^*,\vB) &= -\omega'^*(-\omega,-\vB) = \omega'^*(\omega,\vB),
\end{align*}
we have 
\begin{equation*}
	\PhiT(y'(\omega^*,\vB) ) =  -\PhiT^*(y'(\omega,\vB)),
\end{equation*}
and therefore,
\begin{align*}
	\etaB'_2 (\omega^*,\vB) &= \etaB'^*_2(-\omega,-\vB) = - \etaB'^*_2(\omega,\vB),\ann
	\etaB_2 (\omega^*,\vB) &= \etaB^*_2(-\omega,-\vB) = -\etaB^*_2(\omega,\vB). 
\end{align*}

At the imaginary axis $\omega = \ii \xi$, the conductivity $\etaBT_2(\xi,\vB) = \etaB(\ii\xi,\vB)$ is not real, but it obeys to relations
\begin{equation}\label{eq:cond4}
	\etaBT_2 (\xi,\vB) = \etaBT^*_2 (\xi,-\vB),\ \etaBT_2 (-\xi,\vB) = - \etaBT^*_2 (\xi,\vB).
\end{equation}

The scattering matrix \eqref{eq:r} at the imaginary axes reads
\begin{equation}
	\bm{r}_i(\ii k_E) = - \frac{\xi^2 \etaBT_i + \kB \otimes (\kB\etaBT_i) + \IB \xi k_E \det\etaBT_i}{\xi^2 \tr \etaBT_i + (\kB\kB\etaBT_i) + \xi k_E (1 + \det \etaBT_i)}.
\end{equation}

Due to relations \eqref{eq:cond2} and \eqref{eq:cond4} we obtain 
\begin{align}
	\bm{r}_1(\ii k_E,-\xi) &= \bm{r}_1(\ii k_E,\xi) = \bm{r}^*_1(\ii k_E,\xi)\ann
	\bm{r}_2(\ii k_E,-\xi,\vB) &= \bm{r}^*_1(\ii k_E,\xi,\vB).\label{eq:cond5}
\end{align}

Let us consider the energy \eqref{eq:E2}
\begin{equation}
	\mathcal{E}_2  =  \iint \frac{\dd^2 k}{2(2\pi)^3} \int_{-\infty}^{+\infty} \dd\xi L(\xi,\vB), 
\end{equation}
where
\begin{equation*}
	L (\xi,\vB) = \ln \det  \left[ 1 - e^{ -2 a k_E}  \bm{r}'_1(\ii k_E) \bm{r}_2(\ii k_E) \right].
\end{equation*}
Taking into account the relations \eqref{eq:cond5} we obtain, $L (-\xi,\vB) = L^*(\xi,\vB)$, and therefore,
\begin{align}
	&\mathcal{E}_2 = \iint \frac{\dd^2 k}{2(2\pi)^3} \int_{-\infty}^\infty \dd\xi L (\xi,\vB) \ann
	&= \iint \frac{\dd^2 k}{(2\pi)^3} \int_0^\infty \dd\xi \Re L (\xi,\vB).
\end{align}
Therefore, the $\mathcal{E}_2$ is a real quantity and does not give a contribution to the Casimir friction.

\section{Direct calculation of the boosted PT} \label{Sec:App1}

The PT in the frame $K'$ has the following form 
\begin{equation}
	\tilde{\Pi}'^{ab}(k') = \ii e^2 \iiint \frac{\dd^3 p'}{(2\pi)^3} \tr [\tilde{\gamma}^a\mathcal{S}'(p'+ k')\tilde{\gamma}^b\mathcal{S}'(p')],
\end{equation}
where
\begin{equation}
	\mathcal{S}'(p') = \frac{\tilde{\gamma}^a p'_a + m}{ (\tilde{\gamma}^b p'_b)^2 - m^2 },
\end{equation}
and $\tilde{\gamma}^a = \tilde{\gs}^a_b \gamma^b$, $\tilde{\gs}^a_b = \diag (1,v_F,v_F)$.  

In the frame $K$ one has the same form of PT
\begin{equation}\label{eq:PT}
	\tilde{\Pi}^{ab}(k) = \ii e^2 \iiint \frac{\dd^3 p}{(2\pi)^3} \tr [\tilde{\Gamma}^a\mathcal{S}(p+k)\tilde{\Gamma}^b\mathcal{S}(p)],
\end{equation}
but with matrices $\tilde{\Gamma}^a$ and the spinor Green function $\mathcal{S}$.

To obtain these matrices and Green function we make boost $x^a \to x'^a = \Lambda^{a'}_{\ b} x^b\ (x'=\Lambda x)$. Then the bispinors $\psi(x), \overline{\psi}(x)$ in new coordinates read $\psi'(x') = S \psi(x),\ \overline{\psi}'(x') = \overline{\psi}(x)S^{-1}$ (see, for example, \cite{ClaudeItzykson:2006:qft}). Therefore, 
\begin{equation}
	\tilde{\Gamma}^{a'} = S \tilde{\gamma}^b S^{-1} \Lambda_{\ b}^{a'}, \ \mathcal{S} = \frac{\tilde{\Gamma}^a p_a + m}{ (\tilde{\Gamma}^b p_b)^2 - m^2 }.
\end{equation}
Obviously, that $\Gamma^{a'} = \gamma^{a'}$.   

By straightforward calculations we obtain 
\begin{eqnarray}
	\tilde{\Gamma}^a &=& v_F \gamma^a + u^a (1-v_F) \slashed{u},\ann
	\mathcal{S} &=& \frac{v_F \slashed{p} + (1-v_F)(up) \slashed{u} + m}{p^2v_F^2 + (up)^2 (1 -v_F^2) - m^2},
\end{eqnarray}
where $\slashed{u} = \gamma^a u_a$ and $(up) = u^a p_a$.  Taking into account the above expressions we arrive at the following expression of polarization tensor \eqref{eq:PT}
\begin{equation}
	\tilde{\Pi}^{\mu\nu}(k) = \frac{\ii e^2}{2\pi^3} \iiint  \frac{t^{\mu\nu}\dd^3 p}{Q(q)Q(p)},
\end{equation}
where 
\begin{eqnarray}
	t^{ab}&=& v_F^4 p^{(a}q^{b)} + v_F^2 (1-v_F^2) (uq) p^{(a} u^{b)} + v_F^2 (1-v_F^2) (up) q^{(a} u^{b)} \ann
	&+&  (1-v_F^2) \left[m^2  - (qp) v_F^2 + (up)(uq) (1-v_F^2)\right] u^a u^b\ann
	&+& v_F^2 \left[m^2  - (qp) v_F^2 - (up)(uq) (1-v_F^2) \right] \gs^{ab}.
\end{eqnarray} 
Here, $q = p+k$, $Q(p) = p^2v_F^2 + (up)^2 (1 -v_F^2) - m^2$, and $a^{(ab)} = a^{ab} + a^{ba}$. Then, taking into account the Pauli-Villars renormalization we arrive at  \eqref{eq:PT-2}. 

\section{Zero velocity terms for short distance} \label{Sec:App2}

The direct evaluations integrals in Eqs. \eqref{eq:E0gg}, \eqref{eq:E0gm} where $a\to 0$ has taken, give the following expressions for zero-velocity terms  
\begin{widetext}
	\begin{align}
		\left.\frac{	\mathcal{E}_{\tm,0}^{\gs\gs} }{	\mathcal{E}_{\id} } \right|_{a\to 0} &=   \frac{45}{\pi^4}\frac{\eta _{\gr}^2 }{\eta _{\gr}^2 - v_F^2} \left\{\frac{\eta _{\gr}}{\eta_{\gr} +1} - \frac{v_F^2 \arctanh\left(\frac{\sqrt{1-v_F^2} \sqrt{\eta _{\gr}^2-v_F^2}}{v_F^2+\eta _{\gr}}\right)}{\sqrt{1-v_F^2} \sqrt{\eta _{\gr}^2-v_F^2}}\right\}  = 0.0047, \ann
		\left.\frac{	\mathcal{E}_{\te,0}^{\gs\gs} }{	\mathcal{E}_{\id} }\right|_{a\to 0}  &=    \frac{45}{\pi ^4 \sqrt{1-v_F^2} \eta _\gr \left(1-v_F^2 \eta _\gr^2\right)} \left\{ \frac{2-3 v_F^2 \eta _\gr^2}{\sqrt{1-v_F^2 \eta _\gr^2}} \arctanh \left(\frac{\sqrt{1-v_F^2} \sqrt{1-v_F^2 \eta _\gr^2}}{v_F^2 \eta _\gr + 1}\right) -2\left(1-v_F^2 \eta _\gr^2\right)\arctanh\left(\sqrt{1-v_F^2}\right) \right.\ann
		&\left.+\frac{\sqrt{1-v_F^2} \eta_\gr}{\eta_\gr +1} \left(2 + \eta _\gr -v_F^2\eta_\gr^2 \left(\eta_\gr+1\right) \right)\right\} = 0.000019, 
	\end{align}
for  system of two graphene, and  
	\begin{align}
		\left.\frac{	\mathcal{E}_{\tm,0}^{\gs\m} }{	\mathcal{E}_{\id} }\right|_{a\to 0} &=   \frac{45 \eta _\gr}{\pi ^4} \left\{ \frac{\arctanh \left( \sqrt{1-v_F^2} \right)}{\sqrt{1-v_F^2}} -\frac{\eta_\gr \arctanh \left(\frac{\sqrt{1-v_F^2} \sqrt{\eta_\gr^2 - v_F^2}}{v_F^2 + \eta_\gr} \right)}{\sqrt{1-v_F^2} \sqrt{\eta_\gr^2 - v_F^2}} \right\} = 0.023, \ann
		\left.\frac{	\mathcal{E}_{\te,0}^{\gs\m} }{	\mathcal{E}_{\id} }\right|_{a\to 0}  &=    \frac{45}{\pi ^4 \eta_\gr} \left\{ \eta_\gr - \frac{\arctanh \left(\sqrt{1-v_F^2}\right)}{\sqrt{1-v_F^2}} + \frac{\arctanh \left(\frac{\sqrt{1-v_F^2} \sqrt{1-v_F^2 \eta_\gr^2}}{v_F^2 \eta_\gr + 1}\right)}{\sqrt{1-v_F^2} \sqrt{1-v_F^2 \eta_\gr^2}}\right\}  = 0.0026,
	\end{align}
\end{widetext}
\noindent for graphene/ideal metal configuration. Numerical evaluation has been made for $v_F = 1/300$. For two graphene, the contribution of TE mode is 3 orders smaller than TM mode, while for the second case, one has single order, only. Because $\eta_{\gr}/v_F = 3.4$ we can not expand this expression over $\eta_\gr$.  For $v_F =0$ we obtain 
\begin{equation}
	\left.\frac{	\mathcal{E}_{\tm,0}^{\gs\gs} }{	\mathcal{E}_{\id} }\right|_{a\to 0}  \approx \frac{45 \eta_\gr}{\pi^4}, \ \left.\frac{	\mathcal{E}_{\te,0}^{\gs\gs} }{	\mathcal{E}_{\id} }\right|_{a\to 0}  \approx \frac{15 \eta_\gr^2}{\pi ^4},
\end{equation}
for two graphene, and 
\begin{equation}
	\left.\frac{	\mathcal{E}_{\tm,0}^{\gs\m} }{	\mathcal{E}_{\id} }\right|_{a\to 0}  \approx -\frac{45 \eta_\gr \ln \eta_\gr}{\pi^4}, \ \left.\frac{	\mathcal{E}_{\te,0}^{\gs\m} }{	\mathcal{E}_{\id} } \right|_{a\to 0} \approx \frac{45 \eta_\gr}{2 \pi^4},
\end{equation}
for second case.

%\bibliography{mgr}
%\bibliography{/home/nail/MEGAsync/Bib/my_publ,/home/nail/MEGAsync/Bib/books,/home/nail/MEGAsync/Bib/casimir}
%\bibliographystyle{unsrt}
%\bibliographystyle{apsrev4-2}
\end{document}